\documentclass[conference]{IEEEtran}
\IEEEoverridecommandlockouts

\usepackage{amssymb}
\usepackage{amsmath}

\usepackage[caption=false]{subfig}

\usepackage[labelfont=bf,justification=justified]{caption}
\usepackage{threeparttable}

\usepackage{placeins}

\usepackage{defs}

\usepackage{lineno}

\usepackage{url}

\def\BibTeX{{\rm B\kern-.05em{\sc i\kern-.025em b}\kern-.08em
    T\kern-.1667em\lower.7ex\hbox{E}\kern-.125emX}}

\begin{document}

\title{Tensor-Parallel Emulation of Quantum Circuits \\ with Block-Cyclic Distributed Matrix Product States} 

\author{\IEEEauthorblockN{Jakub Adamski}
\IEEEauthorblockA{\textit{EPCC} \\
\textit{University of Edinburgh}\\
Edinburgh, United Kingdom \\
0009-0000-5221-2070}
\and
\IEEEauthorblockN{Oliver Thomson Brown}
\IEEEauthorblockA{\textit{EPCC} \\
\textit{University of Edinburgh}\\
Edinburgh, United Kingdom \\
0000-0002-5193-8635}
}


\maketitle

\begin{abstract}
Tensor networks establish an adaptable framework for the emulation of quantum circuits. By partitioning exponentially large registers and gates into smaller tensors, this unlocks fast transformations through tensor algebra, and grants fine control over memory, runtime and accuracy. Due to inherently lower spatial footprint, there is a gap in distributed-memory tensor network methods. While certain parallel techniques exist, they are usually limited to direct contraction and sampling problems, and a more general approach is needed for tensor representations like matrix product states~(MPS), which efficiently approximate full quantum state evolution. In this study, we expand the MPS site tensors beyond local memory by introducing a tensor-parallel distribution scheme, where individual dense tensors are evenly scattered across a subset of indices. This is further facilitated by leveraging pivoted QR factorisation instead of slower singular value decomposition~(SVD). We demonstrate the capabilities of our approach by approximately emulating the classically difficult Google's random circuit sampling~(RCS) benchmark. The highest bond dimensions of~$16,384$ is reached, surpassing the accuracy of the state-of-the-art methods by three orders of magnitude on $32$~nodes of ARCHER2. We also show how this helps advance experiments involving more practical quantum phase estimation circuits. Our approach has the potential to enhance numerous algorithms based on dense tensor networks, offering a scalable and naturally load-balanced distribution formula. It is also compatible with other types of parallelism, unlocking new opportunities to push the quantum-classical computational phase boundary. 
\end{abstract}


\section{Introduction}
\label{sec:intro}


\textit{Quantum computing} is an emerging computational paradigm that has been growing in popularity throughout the last decade, as it promises polynomial to exponential speedups when solving certain problems, such as spatial search, order finding, or eigenstate optimisation~\cite{childs-goldstone-04, shor-94, mcclean-romero-16}. While this has been an active research topic for many years, recent developments in hardware and error correction mean that it is a realistic projection that the era of useful quantum devices is due to start in a matter of years~\cite{google-25}. These technological improvements amplify the need for classical emulation, which provides a controlled environment to model and analyse quantum behaviour. However, emulation of quantum circuits is an exponentially hard task -- while a classical register only stores one bitstring at a time, its quantum counterpart can simultaneously hold all possible combinations. Therefore, the access to HPC resources is required to successfully study algorithms designed for modern quantum hardware~\cite{haner-steiger-16}. 

A quantum circuit can be expressed in the language of \textit{tensor networks}, which provides a versatile universal framework for the emulation of quantum systems~\cite{markov-shi-08}. The partitioned initial state is evolved by pairwise contraction (i.e. tensor multiplication) with tensors that realise the components of the emulated circuit. The benefit of this technique is adaptability, as the contractions can follow in any order, optimised either for speed or memory usage~\cite{pan-zhang-22}. In addition, there are other tensor operations that facilitate the representation of quantum information, such as singular value decomposition (SVD), which is adapted to compress/truncate tensor dimensions. Consequently, tensor networks are particularly well-suited for emulation of noisy systems, where accuracy and precision are lower by default~\cite{zhou-stoudenmire-20}. 

However, a major challenge is that larger tensor networks often impose significant memory overheads in the middle of the contraction path. This is common in fields like quantum computing or chemistry, where the state size is exponential in the number of physical constituents (such as particles or qubits). To remedy that, the network can be distributed with a technique known as \textit{index slicing}, converting chosen summations into sum reductions of multiple tensor networks, each of which can be contracted in an embarrassingly parallel manner \cite{solomonik-matthews-14,lyakh-nguyen-22,bayraktar-charara-23}. Another solution is to leverage a structured representation. For instance, matrix product states (MPS) are used to compress the tensors through decomposition and truncation to discard the least important information beyond some fixed dimensions~\cite{fishman-white-22, ibrahim-williams-14}. The trade-off is that decomposition is an expensive operation, and usually creates a major bottleneck that hinders scaling the maximum tensor size, which in turn severely impacts the accuracy that can be achieved. In fact, the SVD scaling is cubic in either of the matrix dimensions, and as it internally relies on matrix multiplication, it is inherently slower than tensor contraction. Additionally, index slicing cannot be used to parallelise decomposition, and hence, the associated implementations suffer from the consequences of Amdahl's law where the serial fraction dominates. 

In this paper, we present \textbf{Quantum Tensor Network Hub (QTNH)} –- a lightweight distributed tensor network library that employs a distribution pattern designed to scatter individual dense tensors across nodes. It offers a simple interface to tensor operations, adopting the tensor structure akin to block-distributed matrices. We leverage this to make the following novel research contributions: 

\begin{itemize}
  \item Introduce a scalable distribution approach to the dense-site MPS evolution algorithm, unlocking unprecedented bond dimension (up to $\chi = 16,384$ on ARCHER2). This is an important milestone in pushing the capabilities of classical emulation, and mapping the quantum-classical computational phase boundary. We show how it applies to both classically hard benchmarks and circuits with practical applications. 
  \item Minimise the decomposition bottleneck by replacing the SVD routine with pivoted QR factorisation\footnote{Note that this is more complex than the standard QR routines based on arbitrary Gram-Schmidt orthogonalisation.}, which achieves better runtime even after compensating for the fidelity loss with larger system size. Although this approach has been suggested before, it was not used due to a major bug in older versions of ScaLAPACK~\cite{wang-hill-17}. 
  \item Demonstrate that our MPI-parallel MPS implementation outperforms the state-of-the-art threaded libraries when emulating the Google's random circuit sampling (RCS) benchmark, not only by allowing scaling into distributed memory, but also when limited to a single node. 
\end{itemize}


The study is organised as follows. In Sec.~\ref{sec:background}, we introduce relevant concepts in the field of quantum computing and tensor networks, while highlighting related work. Sec.~\ref{sec:method} examines the considerations behind QTNH, and sets up circuits to demonstrate the library's capabilities. These are presented and evaluated in Sec.~\ref{sec:results}, which helps pinpoint primary target use-cases, and identify current limitations. Finally, Sec.~\ref{sec:future} enumerates potential improvements and optimisations to our approach.

\section{Background}
\label{sec:background}


In this chapter, we briefly cover the concepts applicable to this work, which comprises basics of quantum computation, emulation and tensor representations\footnote{For more in-depth background, a detailed introduction to quantum computing is provided in~\cite{nielsen-chuang-12}, while tensor networks are elucidated in~\cite{biamonte-19} and~\cite{schollwock-11}.}. 

\subsection{Quantum computing}

\subsubsection{Qubits and statevector}
\label{sec:bkg:qc:qubits}

Similar to digital computing utilising bits, quantum computing is based on quantum bits, called \textit{qubits}. A fundamental difference from their classical counterpart is that a qubit can store a \textit{superposition} of both binary $0$ and $1$ values. To denote this, we use the \textit{Dirac notation}: 

\begin{equation*}
  \ket{\psi} = a\ket{0} + b\ket{1}
\end{equation*}

The complex coefficients $a$ and $b$ are related to the probability of measuring corresponding values by the squared modulus function. This imposes a constraint that $\abs{a}^2 + \abs{b}^2 = 1$. 

Multiple qubits can be combined into a \textit{quantum register} via \textit{tensor product} operation, denoted by a binary operator $\otimes$ (which is often omitted). For instance, two arbitrary qubits are concatenated as follows: 

\begin{equation*}
  \ket{\psi_1} \otimes \ket{\psi_2} = a_1 a_2 \ket{00} + a_1 b_2 \ket{01} + b_1 a_2 \ket{10} + b_1 b_2 \ket{11}
\end{equation*} 

This preserves the probability constraint, i.e. all coefficients squared still add up to one. The notations $\ket{1} \otimes \ket{0} \otimes \ket{1}$, $\ket{1}\ket{0}\ket{1}$ or $\ket{101}$ are all equivalent. Additionally, the decimal integer states $\ket{i}$ can be used in place of the binary format. In general, we write: 

\begin{equation*}
  \ket{\Psi} = \sum_{i=0}^{2^n - 1} \psi_i \ket{i}
\end{equation*}

to represent a register with $n$ qubits. The complex numbers $\psi_i$ form a \textit{statevector} of $2^n$ elements. 

\subsubsection{Gates and entanglement}
\label{sec:bkg:qc:gates}

A quantum state in the register can be modified with \textit{quantum gates}, which describe how new elements are computed based on some linear combination of the old elements. Therefore, a general gate is a $2^n \times 2^n$ matrix. However, in most cases, quantum gates act only on a subset of qubits, effectively taking $2^k \times 2^k$ elements for a $k$-qubit gate. 

Another property of quantum registers is \textit{entanglement}. This occurs when the register can no longer be represented as a tensor product of individual qubits, which can be induced by a two-qubit (or larger) gate. An entangled state instance is the Bell state $\ket{\Phi^+} = \frac{1}{\sqrt{2}} \left(\ket{00} + \ket{11}\right)$. In general, any quantum state can be split into two parts using the following relation: 

\begin{equation*}
  \ket{\psi}=\sum_i \lambda_i \ket{\phi_i^{(A)}} \ket{\phi_i^{(B)}}
\end{equation*}

where $A$ and $B$ label the subsystems. This is called \textit{Schmidt decomposition}, and the subsystems are entangled if there are multiple terms in the summation. Remarkably, this decomposition is equivalent to a singular value decomposition (SVD) of the original state reshaped as a matrix~\cite{nielsen-chuang-12}. 

\subsubsection{Measurement}
\label{sec:bkg:qc:measurement}

A quantum register remains in superposition until it is \textit{measured}. After the measurement, it collapses to the observed value. This also applies to \textit{partial measurements}, where only a subset of qubits are observed. For instance, in a state $\psi_0 \ket{00} + \psi_1 \ket{01} + \psi_2 \ket{10} + \psi_3 \ket{11}$, if the first qubit is measured as $\ket{0}$, the state collapses into $\psi_0' \ket{00} + \psi_1' \ket{01}$, where the coefficients are renormalised so that the probability constraint is preserved: 
\begin{equation*}
  \psi_0' = \frac{\psi_0}{\sqrt{\abs{\psi_0}^2 + \abs{\psi_1}^2}} \quad \text{and} \quad \psi_1' = \frac{\psi_1}{\sqrt{\abs{\psi_0}^2 + \abs{\psi_1}^2}}
\end{equation*}

\subsubsection{Computational phase boundary}
\label{sec:bkg:qc:advantage}

The concepts described in Sec.~\ref{sec:bkg:qc:qubits}--\ref{sec:bkg:qc:measurement} are a basis for novel quantum algorithms that are potentially superior to their classical counterparts. In particular, superposition and entanglement enable massively parallel information processing, but this cannot be leveraged directly because of the need to measure the quantum state. Instead, such parallelism can be used to extract global properties of encoded problems, for example, \textit{quantum phase estimation (QPE)} calculates the eigenvalues of a unitary square matrix~\cite{nielsen-chuang-12}. This algorithm heavily relies on the quantum Fourier transform~(QFT) routine, which we chose as one of the benchmarks in this work. 

Even though the benefits of quantum algorithms have been demonstrated, the major challenge is executing them on hardware. Real qubits and gates tend to be noisy, which limits the available entanglement. Until robust error-correction is achieved, we are in the age of \textit{noisy intermediate-scale quantum (NISQ)} devices. To mark the point when quantum computers can perform some tasks better than classical hardware, the terms such as \textit{quantum supremacy}, \textit{advantage} or \textit{utility} are used~\cite{arute-arya-19, kim-eddins-23}. More generally, for a given problem and computational resources, we can define a \textbf{computational phase boundary} -- which is the scale at which it is optimal to switch to the quantum implementation. 

The first notable instance of quantum supremacy was demonstrated by the Google team in~2019 using the \textit{random circuit sampling (RCS)} algorithm. Its aim is to generate a random distribution that cannot be easily reproduced classically. While the practical applications of this are debatable, the circuit became an important benchmark used to determine the gap between classical emulation and noisy quantum hardware, and hence, it is the second algorithm we employ to stress-test our method~\cite{arute-arya-19, fu-su-24}. 



\subsection{Tensor networks}
\label{sec:bkg:tn}

Tensors are multidimensional arrays, whose \textit{rank} specifies the number of dimensions. They are often visualised as boxes with attached wires, or \textit{tensor indices}. Each index has a corresponding dimension, which is the number of labels it can be assigned. Combinations of labels at each index point to complex \textit{tensor elements}. Therefore, a rank-$r$ tensor with index dimensions $\left( d_0, d_1, \ldots, d_{r-1} \right)$ contains $\prod_{i=0}^{r-1}d_i$ elements. Particularly, rank-$1$ tensors are equivalent to vectors, while rank-$2$ tensors can be treated as matrices. 

\begin{figure}[tb]
  \centering
  \subfloat[]{
    \includegraphics[width=0.42\linewidth]{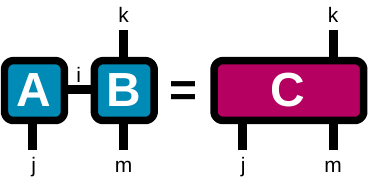}
    \label{fig:tensor-contraction}
  } \hfill
  \subfloat[]{
    \includegraphics[width=0.385\linewidth]{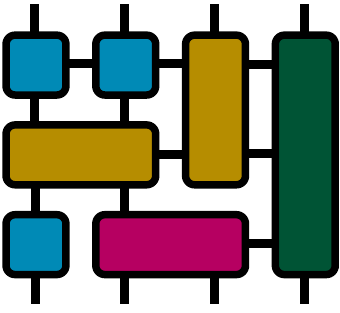}
    \label{fig:example-tn}
  }

  \caption{Graphical representation of tensor networks. \textbf{Fig.~\ref{fig:tensor-contraction}} visualises tensor contraction. \textbf{Fig.~\ref{fig:example-tn}} shows a tensor network with $7$~tensors and $18$~indices ($10$~closed and $8$~open). }
  \label{fig:tn-basics}
\end{figure}

A \textit{tensor contraction} is performed on connected wires, multiplying two tensors similar to matrix multiplication. E.g., two tensors $A_{ij}$ and $B_{klm}$ can be linked on the first and second index respectively, resulting in a tensor $C_{jkm} = \sum_i A_{ij} B_{kim}$ (see Fig.~\ref{fig:tensor-contraction}). Multiple interconnected tensors form a \textbf{tensor network}, which can be contracted by repeatedly multiplying its constituents in a given \textit{contraction order}. In practice, some orders have better computational cost, and finding those poses a challenging optimisation problem. The paired wires are called \textit{closed indices}, while the unpaired ones are \textit{open indices}. An example network is shown on Fig.~\ref{fig:example-tn}. 

\subsubsection{Quantum circuit factorisation}

Any quantum circuit can be regarded as a tensor network -- an $n$-qubit statevector corresponds to a rank-$n$ tensor, while $k$-qubit gates are rank-$2k$. Since usually $n \gg k$, the former tends to be the main memory bottleneck. We can bypass this by factorising the state into smaller tensors using \textit{matrix decomposition} $A = X \cdot Y$ on a reshaped state tensor: 

\begin{equation}
  \Psi^{\sigma_1 \sigma_2 \ldots \sigma_n} = 
    \sum_{i_1} X_{\sigma_1, i_1} Y_{i_1, \sigma_2 \ldots \sigma_n} = 
    \sum_{i_1} M^{\sigma_1}_{i_1} R^{\sigma_2 \ldots \sigma_n}_{i_1}
\end{equation}

The routine above isolates the first site. By iteratively applying this to the remainder tensor $R$, all sites tensors can be computed. It is also valid to perform the splits right-to-left. This procedure generates a construct known as \textbf{matrix product states (MPS)}: 

\begin{equation}
  \Psi^{\sigma_1 \sigma_2 \ldots \sigma_n} = 
    \sum_{i_0 \ldots i_n } M^{\sigma_1}_{i_0,i_1} M^{\sigma_2}_{i_1,i_2} \ldots 
    M^{\sigma_n}_{i_{n-1},i_n}
\end{equation}

Although the decomposition scheme is arbitrary, the commonly used algorithms are \textbf{QR/LQ factorisation} and \textbf{singular value decomposition (SVD)}, since they involve unitary matrices that result in a \textit{canonical} MPS. This helps simplify computations and optimises entanglement distribution. In addition, the SVD and QR factorisation with pivoting (which is more robust than the standard QR method) allow efficient compression of site tensors through \textit{truncation}~\cite{schollwock-11}.

\subsubsection{Distributed tensor algebra}

While tensor networks were originally intended to keep large problems entirely within a single machine's memory, as the problem size and entanglement unavoidably soar, there is a growing interest in parallel techniques that can be implemented on HPC clusters. We identified three types of parallelism available to distribute tensor networks: 

\begin{itemize}
  \item \textbf{Slice parallelism} -- factors out selected contractions, converting them to sum reductions of multiple subnetworks, known as \textit{slices}. The slices are embarrassingly parallel to contract, followed by a global reduction. 
  
  \item \textbf{Tensor parallelism} -- scatters the elements of individual tensors. High-performance linear algebra libraries can be used to operate on distributed tensors reshaped as matrices. 
  
  \item \textbf{Network parallelism} -- when multiple operations involve disjoint sets of  tensors, they can be broadcast to independent ranks. This uncouples the computations and lets them be executed in parallel. 
\end{itemize}

\subsection{Prior work}

The majority of existing tensor network circuit emulation work relies on sampling -- an embarrassingly-parallel direct approach that computes individual amplitudes~\cite{fu-su-24}. By contrast, we adopt the MPS evolution method, which retains the entire state, albeit often approximated~\cite{schieffer-markidis-25}. However, we found there is a clear lack of tools that realise this method in distributed memory, with most parallel implementations limited to a threaded BLAS backend. This hinders the scalability of state tensors, which is crucial to maximise emulation fidelity~\cite{zhou-stoudenmire-20}. We designed QTNH aiming to fill this gap. 

It is worth noting that some prior studies do employ a distributed MPS implementation, but within the context of the \textit{density matrix renormalisation group (DMRG)} ground state search algorithm~\cite{schollwock-11}. This is a distinct method for general quantum systems that often relies on block-sparse tensors, which are significantly easier to scale up. For instance, a DMRG with bond dimension of $\chi = 65,536$ was attained on Google's tensor processing units when leveraging FP32 tensor elements and hard-drive storage~\cite{ganahl-beall-23}. Similarly, another study used the CTF library to benchmark block-sparse tensor-parallel DMRG with $\chi$ up to $32,768$~\cite{levy-solomonik-20}. 




\subsubsection{Parallel tensor libraries}

We summarise the related software projects in Table~\ref{tab:software}. \textit{ITensor} and \textit{quimb} are state-of-the-art libraries used for MPS algorithms, but the performance of their dense tensor operations is barred by the underlying BLAS routines~\cite{fishman-white-22, gray-18}. NVIDIA's \textit{cuQuantum} ecosystem with \textit{cuTensorNet} proprietary module also has an MPS interface, and is designed to offload the computations to a GPU, however, as of the writing of this, it only supports a single-GPU MPS backend~\cite{bayraktar-charara-23}. Finally, the most closely related prior project is \textit{Cyclops Tensor Framework (CTF)}, as it too involves tensor-level parallelism and ScaLAPACK for distributed linear algebra~\cite{solomonik-matthews-14}. The main difference from QTNH is that CTF's target use-case is quantum chemistry, and hence, its data representation and computations are optimised for sparse tensors. In addition, it does not handle truncating decomposition routines beyond the SVD, such as the pivoted QR factorisation, which is a major focus in this study. As CTF has a much lower-level interface than the other mentioned libraries, it is rarely used for the emulation of quantum circuits. 

\begin{table*}[tb]
\centering
\begin{threeparttable}
\caption{Software libraries comparison}
\begin{tabular}{l|l|l|l}
  \textbf{Library} & \textbf{Language} & \textbf{Parallelism} & \textbf{Other notes} \\
  \hline\hline
  ITensor~\cite{fishman-white-22} & Julia & BLAS & GPU backend \\
  \hline
  quimb~\cite{gray-18} & Python & BLAS & \begin{tabular}[c]{@{}l@{}}
    CUDA backend, PETSc \\ matrix interface
  \end{tabular} \\
  \hline
  cuQuantum~\cite{bayraktar-charara-23} & C++ & slice, network & \begin{tabular}[c]{@{}l@{}}
    NVIDIA GPU backend, \\ proprietary modules
  \end{tabular} \\
  \hline
  ExaTN~\cite{lyakh-nguyen-22} & C++ & slice, network & \begin{tabular}[c]{@{}l@{}}
    complex dependencies, \\ discontinued support
  \end{tabular} \\
  \hline
  CTF~\cite{solomonik-matthews-14} & C++ & tensor & 
  \begin{tabular}[c]{@{}l@{}}
    sparse tensor focus, \\ low-level interface
  \end{tabular}
\end{tabular}
\label{tab:software}
\end{threeparttable}
\end{table*}

\section{Method}
\label{sec:method}

\textbf{Quantum Tensor Network Hub (QTNH)} is a portable tensor network library made for manipulating distributed dense tensors using our novel distribution pattern, which combines both tensor and network types of parallelism. Moreover, we introduce a procedure to efficiently extend the distributed matrix routines from ScaLAPACK to tensors. As a result, the performance of our library is directly linked to the complexity of the underpinning linear algebra methods, which in ScaLAPACK are optimised for distributed computing~\cite{blackford-choi-97}. In particular, we show that for large dense tensors associated with MPS evolution, the time spent permuting the tensors into matrices is negligible compared to the linear algebra calls. 




\subsection{Tensors}
\label{sec:method:tensors}

\subsubsection{Element layout and distribution}
\label{sec:method:tensors:layout}

Dense tensors are challenging to address efficiently, due to effectively random memory access pattern. A common technique, which we also employ in QTNH, is a \textit{flat contiguous layout} with \textit{row-major (lexicographic) ordering}. This means that contiguous accesses only occur for the rightmost index. 

The unique feature of QTNH is the distribution pattern. We \textbf{scatter} a rank-$n$ tensor between multiple MPI ranks by fixing its leftmost indices. The first $n_d$ indices label the processes that hold the tensor, while the remaining $n_l = n-n_d$ indices address the local elements -- we call them \textit{distributed} and \textit{local indices} respectively. As an example, a rank-$4$ tensor with shape $(2, 3; 4, 2)$ and $n_d=2$ has a distributed size of~6 and local size of~8\footnote{The semicolon is a shorthand for the distributed/local index split}. 

\begin{figure*}[tb]
  \centering
  \includegraphics[width=\linewidth]{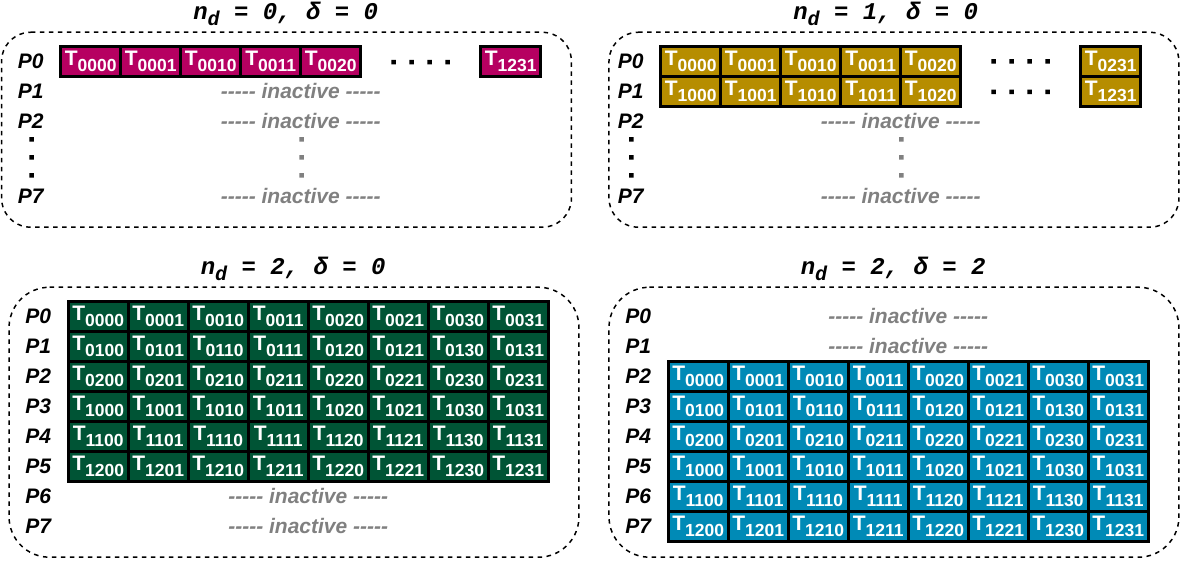}
  \caption{Tensor distribution in QTNH of a $\left(2,3,4,2\right)$ tensor. }
  \label{fig:qtnh-distribution}
\end{figure*}

Furthermore, tensors can also be \textbf{offset} by~$\delta$ relative to rank~0, which shifts the process label specified by the distributed indices. The ranks that don't contain any tensor elements are marked as \textit{inactive}. This allows uncoupling distant tensors. As a result, our distribution enables parallelism at both tensor and network levels. This is visualised in Fig.~\ref{fig:qtnh-distribution}. 



\subsubsection{Index permutation}
\label{sec:method:tensors:permutation}

Index permutation/transposition rearranges the elements such that they are addressed with a permutation of the original indices. The new ordering is based on a \textit{permutation tuple} that has one of the following two types:

\begin{itemize}
  \item Source tuple $s(i): P_s(k_i) = k_{s(i)}, T'(k_i) = T(P_s(k_i))$ 
  \item Target tuple $t(i): P_t(k_{t(i)}) = k_i, T'(P_t(k_i)) = T(k_i)$ 
\end{itemize}

\begin{figure*}[tb]
  \centering
  \includegraphics[width=\linewidth]{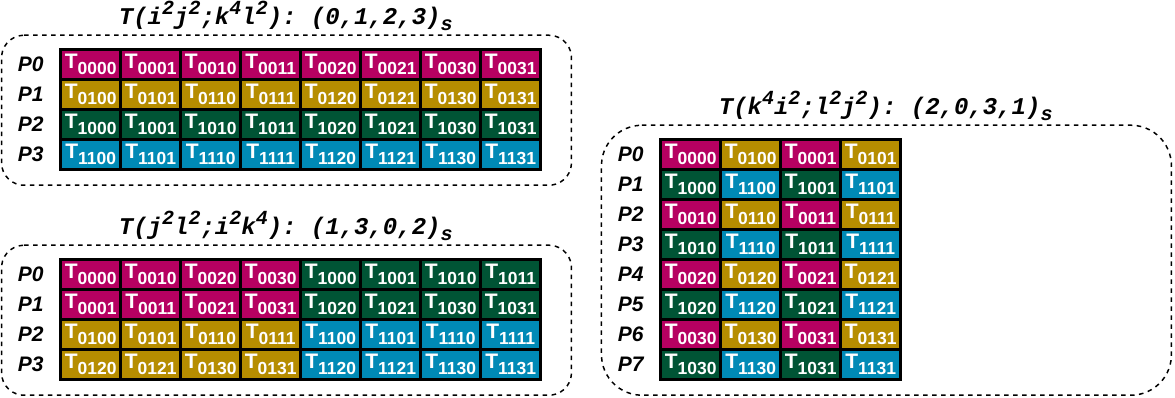}
  \caption{Examples of permuting indices of a $\left(2;2,4,2\right)$ tensor. Moving the 4-dimensional local index to the distributed position is an \textit{asymmetric permutation} and requires more MPI ranks than the original tensor. }
  \label{fig:idx-perm}
\end{figure*}

In a general case, this routine involves asymmetric all-to-all communication, where the numbers of senders and receivers might differ, as shown in Fig.~\ref{fig:idx-perm}. In QTNH, this is handled with a call to \texttt{MPI\_Alltoallv}, which communicates the elements arranged in a nested contiguous-resized MPI datatype on both ends. Our approach supports automatic merging of type initialisations where possible, which is important to address limitation in some MPI communication libraries. 


The approach above is universal, but \texttt{MPI\_Alltoallv} is known to have a high communication cost. While, this was not apparent in our experiments, we also introduced an optimised communication-free algorithm for local transposition. It is not general, and can only be applied when the distributed indices are fixed, but this is a common case for MPS tensors. 



\subsubsection{Block-cyclic matrices}
\label{sec:method:tensors:matrices}

To offload tensor operations to ScaLAPACK, which uses the \textit{block-cyclic distribution pattern}, we introduce direct conversion between QTNH tensors and ScaLAPACK matrices. Conveniently, this solely relies on index permutation. We first divide the indices into six groups, defined in Table~\ref{tab:bc-indices}. 

\begin{table}[tb]
\centering
\begin{threeparttable}
\caption{Tensor index groups mapping to block-cyclic matrices}
\begin{tabular}{l|K{2cm}|K{2cm}|K{2cm}}
        & Cyclic                & Distributed           & Block                 \\
\hline
Row     & $\left\{i_c\right\}$  & $\left\{i_d\right\}$  & $\left\{i_b\right\}$  \\
Column  & $\left\{j_c\right\}$  & $\left\{j_d\right\}$  & $\left\{j_b\right\}$
\end{tabular}
\label{tab:bc-indices}
\end{threeparttable}
\end{table}


The assignment of groups depends on the offloaded operation, but the split between distributed and local indices should be fixed to limit communication. The tensor is permuted based on the following tuple: 

\begin{equation}
  \label{eqn:ptup}
  P_s = \left(\{i_d\}\{j_d\};\{j_c\}\{j_b\}\{i_c\}\{i_b\}\right)
\end{equation}

Local rows and columns are reversed due to Fortran's column-major indexing. After a ScaLAPACK routine is called, the data is permuted back to the QTNH format. A further optimisation would be to permute the tensors lazily, but this was deemed to be beyond the scope of this work, as permutations in our experiments are proportionately cheap. 


\subsection{Tensor networks}
\label{sec:method:tn}

To characterise contractions, we use \textbf{bonds}, which connect two tensors via a set of wires. A unique restriction in QTNH is that indices in a wire need to be of the same dimensions and type (i.e.\ distributed/local), to simplify the alignment and sum reduction of contracted tensors. A \textbf{tensor network} is defined by a collection of tensors and bonds. 



\subsubsection{Bond contraction}
\label{sec:method:tn:contraction}

Two tensors can be contracted in line with the associated bond by summing products of elements with the same joint indices. QTNH offloads this to ScaLAPACK's \texttt{PZGEMM} after permuting the tensors. The closed indices become matrix columns, and open indices turn into rows, with the second matrix marked as transposed. Tracing over wires of the same tensor is achieved by contracting two wires with both ends of an \textit{identity tensor}. 



\subsubsection{Tensor decomposition}
\label{sec:method:tn:decomposition}

To decompose a tensor, we use one of the matrix decomposition routines from Table~\ref{tab:decomposition}. Block-cyclic permutation might be preferred for performance reasons. In addition, some methods like the SVD require blocks to be square. Therefore, it is left up to the user to assign indices to sets from Eqn.~\ref{eqn:ptup}. 

\begin{table}[tb]
\centering
\begin{threeparttable}
\caption{Tensor decomposition routines}
\begin{tabular}{l|l|l}
Decomposition     & Formula            & ScaLAPACK routine(s)                 \\
\hline
SVD               & $M = USV^\dagger$  & \texttt{PZGESVD}                     \\
LQ                & $M = LQ$           & \texttt{PZGELQF} + \texttt{PZUNGLQ}  \\
QR                & $M = QR$           & \texttt{PZGEQRF} + \texttt{PZUNGQR}  \\
QR with pivoting  & $M = QRP^T$        & \texttt{PZGEQRP} + \texttt{PZUNGQR}

\end{tabular}
\label{tab:decomposition}
\end{threeparttable}
\end{table}

\subsection{Block-cyclic distributed MPS}
\label{sec:method:bcmps}

\begin{figure}[tb]
  \centering
  \subfloat[]{
    \includegraphics[width=0.25\linewidth]{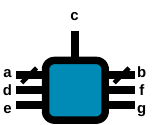}
    \label{fig:bcmpsc:site}
  } \hfill
  \subfloat[]{
    \includegraphics[width=0.65\linewidth]{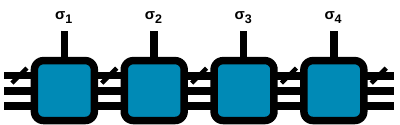}
    \label{fig:bcmpsc:mps}
  }
  
  \caption{Block-cyclic MPS tensor diagram. \textbf{Fig.~\ref{fig:bcmpsc:site}} labels the ordered indices of a site tensor, while \textbf{Fig.~\ref{fig:bcmpsc:mps}} presents how the sites are interconnected.}
  \label{fig:bcmpsc}
\end{figure}

We can leverage the distribution formula of QTNH to construct an MPS tensor network where site tensors are compatible with ScaLAPACK block-cyclic matrices. This is achieved by using rank-$7$ tensors for each site, as shown on Fig.~\ref{fig:bcmpsc:site}, corresponding to $T_{abcdefg}$ tensors with the following shape:

\begin{equation}
  \left(\chi_d, \chi_d; D, \chi_c, \chi_b, \chi_c, \chi_b\right)
\end{equation}

It is straightforward to convert this into a block-cyclic matrix, by treating $a,cd,e$ as row indices, and $b,f,g$ as column indices (distributed, cyclic and blocking respectively). Note that $c$ and $d$ are both cyclic row indices. The only associated permutation costs are due to conversion between C and Fortran indexing, which is an entirely local operation. We fix the site dimensions at the beginning of the program, as they usually quickly become saturated. 

\subsubsection{Canonicalisation} 

MPS can be left- or right-canonicalised up to a chosen site. These are implemented using QR and LQ decomposition routines, respectively, which don't require truncation. All canonical sites are unitary. 

\subsubsection{Tensor application}

When a tensor is applied to a subset of sites in the MPS, it is contracted together with all the site tensors between the minimum and maximum target qubit. Afterwards, the MPS is retrieved by performing a truncated decomposition (SVD or pivoted QR) for each contracted site. This is usually only used for small ranges of qubits, e.g. nearest-neighbour interactions. 

\subsubsection{MPO application}

The \textbf{matrix product operators (MPO)} can model longer-range interaction without the need to construct and decompose massive matrices. In QTNH, MPOs follow a standard implementation with no distribution, since their bond dimensions $m$ are expected to be small. The MPS and MPO is contracted using a zip-up algorithm~\cite{schollwock-11, stoudenmire-white-10}. 

\subsubsection{Site permutation}

Instead of applying large MPOs, sometimes it is faster to reorder the MPS sites through a sequence of nearest-neighbour swaps. These are achieved by contracting the site tensors, permuting their physical indices, and decomposing back, which potentially involves truncation. To realise a given site permutation, we use an algorithm based on the bubble sort, where we rearrange the target permutation tuple as we apply the swaps. 








\subsection{Quantum circuits}

To validate our MPS distribution pattern, we selected two circuits -- the \textbf{inverse quantum Fourier transform~(IQFT)} and Google's \textbf{random circuit sampling~(RCS)}~\cite{nielsen-chuang-12, arute-arya-19}. They help juxtapose two contrasting emulation scenarios, where the IQFT transforms the existing entanglement, while the RCS maximises new entanglement. Here, we share the implementation details and considerations for both circuits. We also introduce a new approximate fidelity metric, efficient to obtain with the MPS representation. 

\subsubsection{Inverse quantum Fourier transform} 
\begin{figure*}[tb]
  \centering
  \subfloat[]{
    \includegraphics[width=0.35\linewidth]{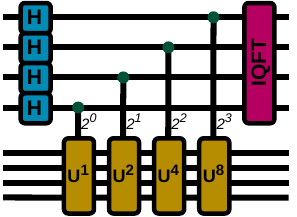}
    \label{fig:iqft:qpe}
  } \hfill
  \subfloat[]{
    \includegraphics[width=0.55\linewidth]{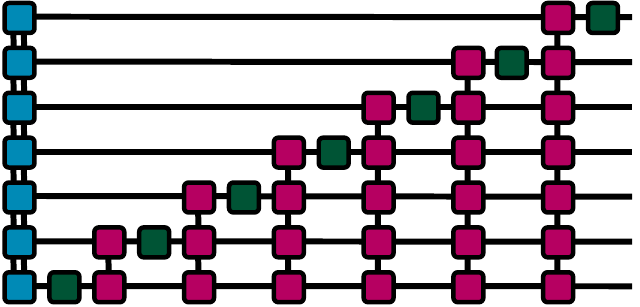}
    \label{fig:iqft:mps}
  }
  
  \caption{Inverse Quantum Fourier Transform. \textbf{Fig.~\ref{fig:iqft:qpe}} presents the application of IQFT in $8$-qubit phase estimation, while \textbf{Fig.~\ref{fig:iqft:mps}} shows a $7$-qubit IQFT structure as a tensor network of Hadamard gates and controlled phase MPOs.}
  \label{fig:iqft}
\end{figure*}

While quantum Fourier transform is not an entangling circuit in itself, its inverse is a common component of larger circuits, such as \textit{quantum phase estimation (QPE)} (Fig.~\ref{fig:iqft:qpe}). As it appears in the last phase of the algorithm, we can assume that its input is already highly entangled, and the IQFT transforms it into an interpretable result. We can emulate this by fixing the saturation of the initial bond dimensions. 

The IQFT can be implemented as a sequence of Hadamard gates, and long-range MPOs with $m=2$, which combine the controlled phase gates with the same control qubit~\cite{chen-stoudenmire-23}. The MPO structure is shown below. 

\begin{equation}
  \begin{pmatrix}
    \hat{\Pi}_0 & \hat{\Pi_1}
  \end{pmatrix} 
  \begin{pmatrix}
    \hat{I} & 0 \\
    0 & \hat{P}_2
  \end{pmatrix}
  \cdots
  \begin{pmatrix}
    \hat{I} & 0 \\
    0 & \hat{P}_{k-1}
  \end{pmatrix}
  \begin{pmatrix}
    \hat{I} \\
    \hat{P}_k
  \end{pmatrix}
\end{equation}

$\hat{\Pi}_k$ is a projector onto state $\ket{k}$, while $\hat{P}_k$ represents phase shift by $\frac{2\pi}{2^k}$. The resultant IQFT circuit is displayed in Fig.~\ref{fig:iqft:mps}. The input state to the IQFT can be partially saturated to a fraction of $\pi \leq 1.0$ by randomly filling the first $\pi\chi \times \pi\chi$ elements of the MPS sites for each physical dimension. This is nearly always guaranteed to produce matrices of required rank~\cite{feng-zhang-07}. 

\subsubsection{Random circuit sampling}

The random sampling was originally used by Google to demonstrate quantum supremacy~\cite{arute-arya-19}. Here, we reproduce the RCS circuit on Sycamore topology with $53$~active qubits and $20$~layers that follow the $ABCDCDAB$ pattern. The entangling $\fSim$ gates have the following form in our implementation:

\begin{equation}
\label{eqn:fsim}
  \hat{\fSim} = \begin{pmatrix}
    1 &  0 &  0 & 0 \\
    0 &  0 & -i & 0 \\
    0 & -i &  0 & 0 \\
    0 &  0 &  0 & e^{-i\pi/6}
  \end{pmatrix}
\end{equation}

The main goal of this circuit is to quickly generate high levels of entanglement, which makes emulating it an especially challenging task. While we won't be able to achieve fidelities higher than Google's hardware ($\mathcal{F}=0.002$) due to the limitations of MPS with single-qubit sites, we show that a significant improvement over the state-of-the-art libraries can be accomplished when the site tensors are scaled beyond local memory. 

One of the challenges when emulating the RCS circuit is its 2-dimensional connectivity. Since the MPS representation comprises a linear chain of tensors, an embedding is necessary, which makes one type of the interactions (horizontal/vertical) effectively long-range. Just like with the QFT, we can construct MPOs of the entangling gates between sites $i$ and $i+k$ of the form shown below. $\hat{\Pi}_{0}$ and $\hat{\Pi}_{0}$ are projectors, while $\hat{S}_{-}$, $\hat{S}_{+}$ are lowering/raising operators, i.e. $\ket{0}\bra{1}$, $\ket{1}\bra{0}$ respectively. The corresponding MPO has a bond dimension of~$4$. 


\begin{equation}
  \begin{pmatrix}
    \hat{\Pi}_0 \\ 
    -i\hat{S}_{-} \\ 
    -i\hat{S}_{+} \\ 
    e^{i\pi/6}\hat{\Pi}_1
  \end{pmatrix} ^T
  \begin{pmatrix}
    \hat{I} & 0 & 0 & 0 \\
    0 & \hat{I} & 0 & 0 \\
    0 & 0 & \hat{I} & 0\\
    0 & 0 & 0 & \hat{I}
  \end{pmatrix}^{k-1}
  \begin{pmatrix}
    \hat{\Pi}_0 \\ 
    \hat{S}_{+} \\ 
    \hat{S}_{-} \\ 
    \hat{\Pi}_1
  \end{pmatrix} 
\end{equation}

Unlike the QFT, where multiple phase gates can be combined into a single MPO, each long-range $\fSim$ gate needs a new MPO. This is not an efficient approach, so in an attempt to improve it, we introduce an alternative method, which relies on permuting the MPS between embeddings with horizontal and vertical nearest neighbours. Hence, we need to insert an additional site permutation procedure between the $AB$ and $CD$ RCS layers. The procedure is based on the bubble sort algorithm, and repeatedly swaps the neighbouring sites, until the desired embedding is reached. For the RCS emulation this ends up requiring exactly $397$~swaps in either direction. 

\begin{figure*}[tb]
  \centering
  \subfloat[]{
    \includegraphics[width=0.55\linewidth]{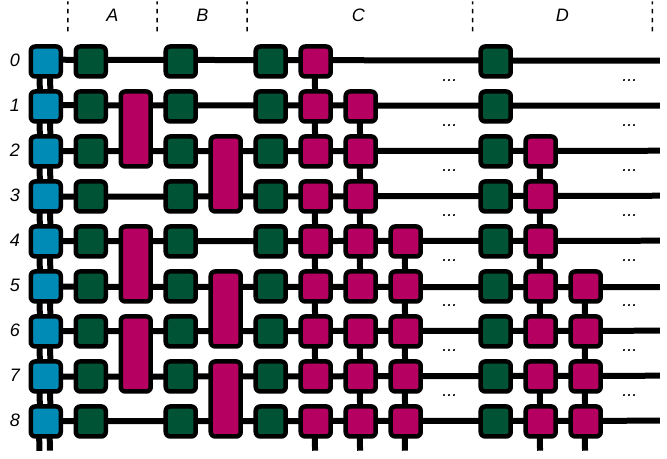}
    \label{fig:rcs:mpo}
  } \hfill
  \subfloat[]{
    \includegraphics[width=0.40\linewidth]{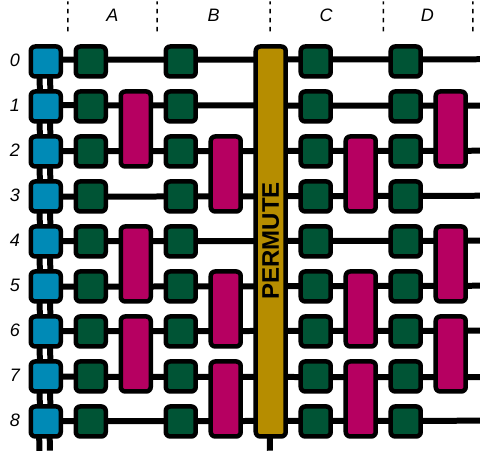}
    \label{fig:rcs:swap}
  }
  
  \caption{First $8$~qubits and $4$~layers of the RCS circuit implemented based on Sycamore connectivity using two approaches. \textbf{Fig.~\ref{fig:rcs:mpo}} handles long-range interactions as MPOs, while \textbf{Fig.~\ref{fig:rcs:swap}} introduces a permutation layer to bring the interacting qubits together.}
  \label{fig:rcs}
\end{figure*}

Both the MPO and permutation approaches are visualised in Fig.~\ref{fig:rcs}. We also share further details on our implementations of the quantum circuits and hardware topology embedding in~\ref{sec:app:qc}. 


\subsection{Circuit fidelity}

To reliably measure the fidelity, we need an overlap between the exact output state $\ket{\Psi_E}$ and the truncated state $\ket{\Psi_T}$: 

\begin{equation}
  \mathcal{F} = \abs{\braket{\Psi_E}{\Psi_T}}^2
\end{equation}

However, it is infeasible to calculate the exact state for high-entanglement circuits. We instead employ a metric called \textbf{norm fidelity}, defined as: 

\begin{equation}
  \bar{\mathcal{F}} = \norm{\ket{\Psi_T}}^2 = \braket{\Psi_T}{\Psi_T}
\end{equation}

\begin{figure*}[tbp]
  \centering
  \includegraphics[width=.8\linewidth]{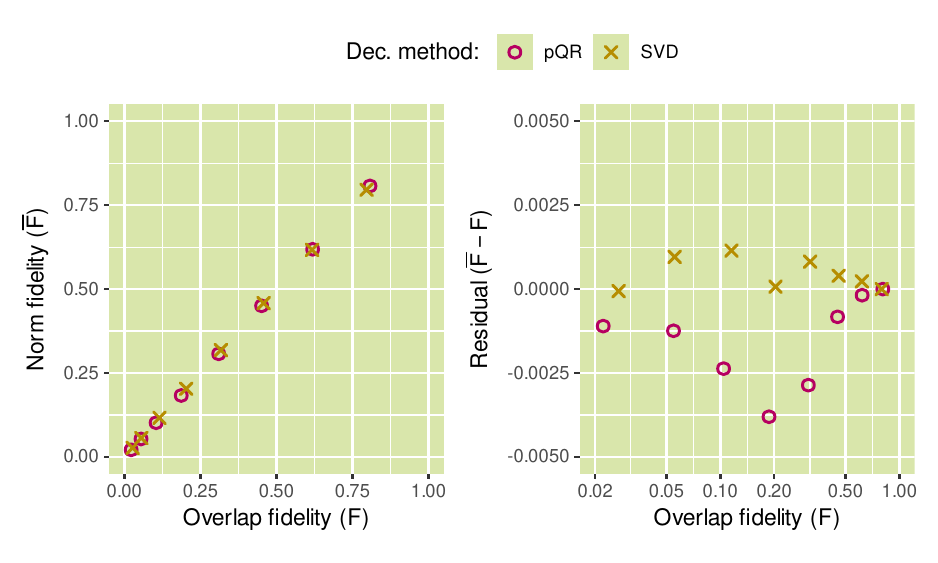}
  \caption{Empirical demonstration of equivalence between standard fidelity and norm fidelity for both SVD and pivoted QR decomposition. The plot on the left shows that the points lie on a $y = x$ line, while the one on the right displays the corresponding residuals ($r = \bar{\mathcal{F}} - \mathcal{F}$). Given that there is no clear asymptotic behaviour in the latter, it is likely that the norm metric becomes less accurate at tiny fidelities. The QR method appears to mostly underestimate, while the SVD tends to overestimate the fidelity. }
  \label{fig:fidelity-overlap}
\end{figure*}

This is empirically demonstrated to be approximately equal to the overlap fidelity metric, as shown in Fig.~\ref{fig:fidelity-overlap}. We can also define the average norm fidelity per interaction as $\bar{f}_{av} \approx \bar{\mathcal{F}}^{1/n_2}$, where $n_2$ is the number of 2-qubit gates. Similarly, the average norm fidelity per swap is $\bar{f}_{swap} \approx \left(\bar{\mathcal{F}'}/\bar{\mathcal{F}}\right)^{1/n_s}$, such that $\bar{\mathcal{F}}$ and $\bar{\mathcal{F'}}$ are norm fidelities before/after the permutation layer, and $n_s$ is the number of swaps. 

\section{Results}
\label{sec:results}


We run experiments on ARCHER2 on up to $32$~standard nodes ($128$~cores and $256$~GB RAM) in the rank-per-core configuration. The parameters of the target circuits were selected such as to reflect the limitations in the state of the art: 

\begin{enumerate}
  \item A 100-qubit IQFT circuit, which is way beyond the limits of statevector simulation and demonstrates the benefits of tensor networks. We also take advantage of large bond dimensions that are difficult to compute using non-distributed libraries. 
  \item A full $53$-qubit and $20$-layer RCS circuit used by Google to demonstrate quantum supremacy. Although this is known to be impractical to emulate using a na\"ive MPS algorithm, it provides a good framework for comparison of tensor network libraries. 
\end{enumerate}

In the first two sections, we use a downscaled versions of the experiments above to identify and profile the most optimal emulation approach. Then, we compare our best configuration against state-of-the-art libraries for MPS emulation – ITensor and quimb. Finally, we show how our approach can be used to push the limits of circuit emulation for both moderately and highly entangled circuits. 




\subsection{Experimental setup}
\label{sec:results:setup}

In Sec.~\ref{sec:method:tn}--\ref{sec:method:bcmps}, we discussed various implementations of tensor decomposition and long-range interactions. In addition, the decomposition routines can further benefit from block-cyclic decomposition, where the block size is fine-tuned for better performance. In this section, we benchmark the simplified versions of our quantum circuits to determine the optimal parameters for the best method to be used in large-scale experiments. 


\subsubsection{Bond truncation}

We find that one of the most effective improvements to current MPS methods is to change the decomposition method from SVD to pivoted QR factorisation. This has been suggested before, but was not employed due to a major bug in the ScaLAPACK implementation at the time~\cite{wang-hill-17}. However, the problem appears to have been fixed in a recent release, with the QR algorithm itself further improved~\cite{bujanovic-drmac-19}. We compare both decomposition approaches in QTNH, and find that the latter is considerably faster, at a minor loss in fidelity. Fig.~\ref{fig:rcs-setup} plots fidelity against runtime of a number of RCS experiments, and it is clear that much better fidelities can be achieved faster with the QR method, even though they usually require larger overall bond dimensions. In fact, thanks to this technique, we managed to speed up some emulation instances by factors of~2 and above. 


\subsubsection{Long-range interactions}

\begin{figure*}[tbp]
  \centering
  \includegraphics[width=.8\linewidth]{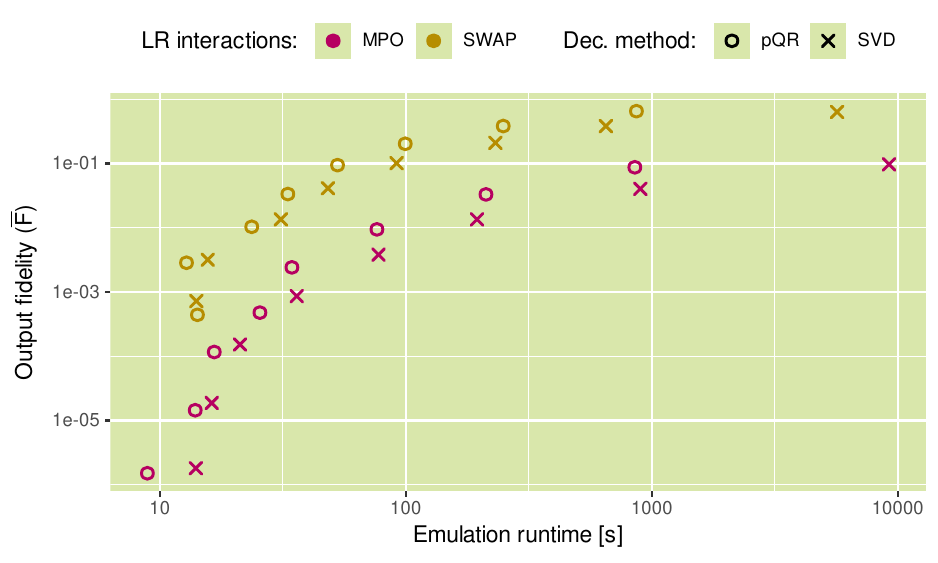}
  \caption{Comparison of RCS emulation approaches, considering decomposition and long-range interactions. Higher fidelity values are desired, while avoiding longer runtimes. SWAP + QR appears to offer the best trade-off for the circuits benchmarked.}
  \label{fig:rcs-setup}
\end{figure*}

Also in Fig.~\ref{fig:rcs-setup}, we contrast the discussed implementations of long-range interactions in QTNH. Only the RCS circuit is considered, as the IQFT naturally benefits from the MPO approach by combining multiple controlled gates into a single operator. However, it is clear that the SWAP-based method performs significantly better for the RCS. 


\subsubsection{Block-cyclic representation}

Although ScaLAPACK is based on block-cyclic matrices, this representation does not offer any benefits over block decomposition for multiplication of evenly distributed matrices prevalent in QTNH. However, this changes for decomposition, where more sophisticated load-balancing is needed. In our approach, the blocking size is controlled by the block dimension $\chi_b$. 

\begin{figure*}[tbp]
  \centering
  \includegraphics[width=.8\linewidth]{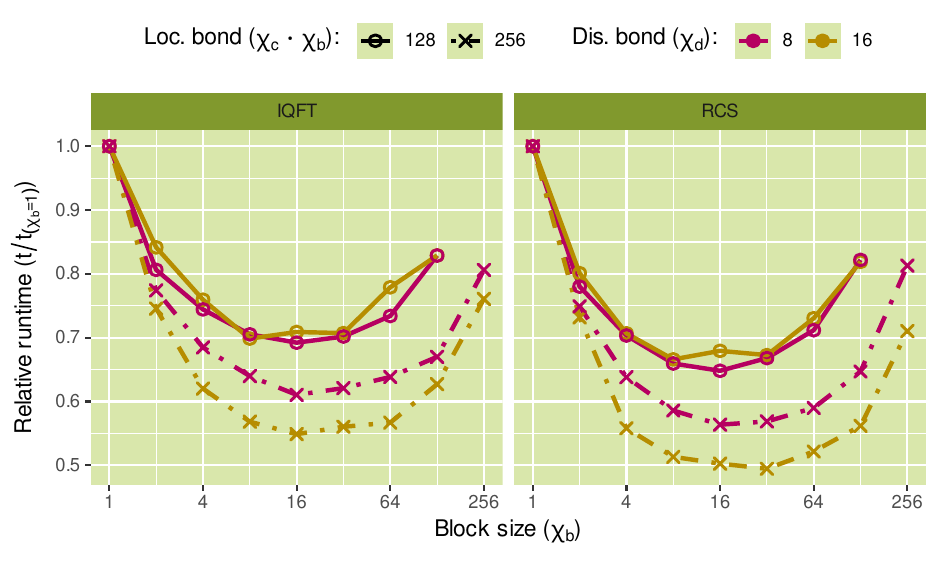}
  \caption{Block size comparison for IQFT~(left) and RCS~(right) circuits. Runtimes are normalised relative to $\chi_b = 1$, and smaller values are better. By choosing $\chi_b$ from $8$ to $32$, we can trim off nearly half of the runtime compared to purely cyclic decomposition.}
  \label{fig:block-size}
\end{figure*}

We investigate the optimal size of $\chi_b$ in decomposition routines for both IQFT and RCS circuits. For a given experiment, the total bond dimension ($\chi$) is kept constant, while we vary the splits between block and cyclic indices, from pure cyclic to pure block variants. Fig.~\ref{fig:block-size} demonstrates that the best block sizes lie between $8$ and $32$, and hence, we settle at fixing $\chi_b = 16$. 

\subsection{Profiling}

Here, we analyse the performance of a 4-layer partial RCS circuit using the parameters determined in Sec.~\ref{sec:results:setup}. This setup takes $\leq \frac{1}{5}$ of runtime while covering all routines in similar proportions, as it is applied cyclically in the full circuit. To include internode communication, two saturated nodes of ARCHER2 are used ($256$~ranks). We choose three local array sizes, and since the block size is fixed at $\chi_b = 16$, we only vary $\chi_c = 8/16/32$, which corresponds to bond dimensions of $\chi = 2048/4096/8192$, respectively. The profiles are generated with the Linaro Forge Arm MAP profiler. 

\begin{figure*}[!htp]
  \centering
  \subfloat[]{
    \includegraphics[width=.9\linewidth]{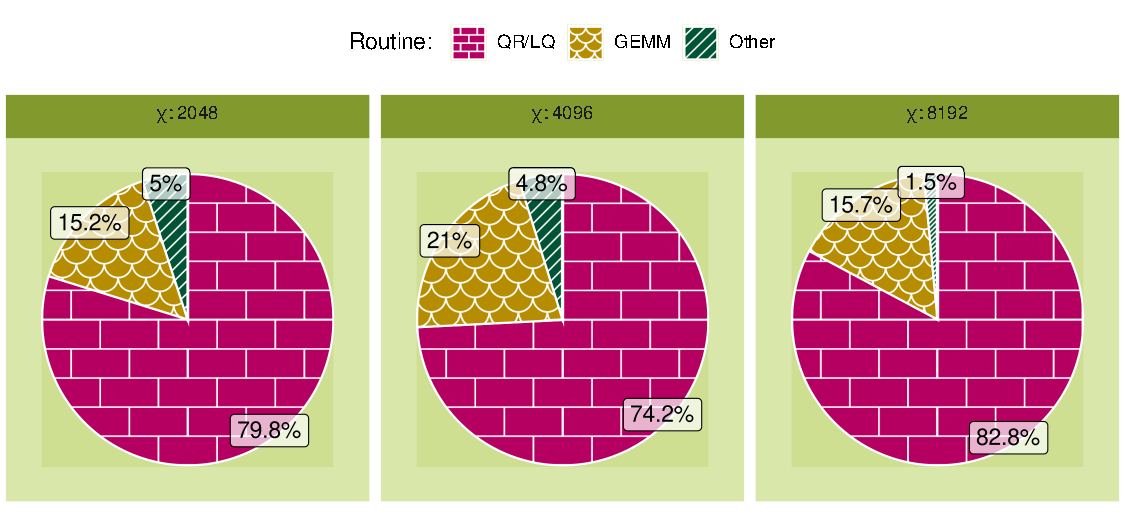}
    \label{fig:profiles:la}
  } \\
  \subfloat[]{
    \includegraphics[width=.9\linewidth]{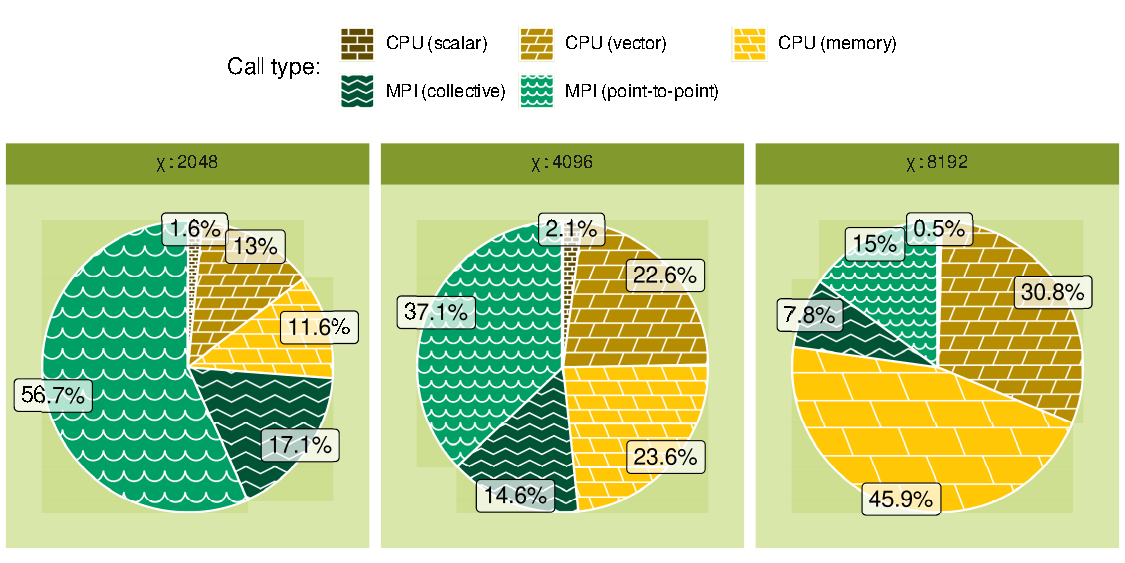}
    \label{fig:profiles:op}
  }
  \caption{Runtime profiles of three partial RCS experiments at different local array sizes. \textbf{Fig.~\ref{fig:profiles:la}} shows the split between decomposition, matrix multiplication and other function calls, where the former always dominates. \textbf{Fig.~\ref{fig:profiles:op}} investigates how the program changes from communication-bound to compute-bound at different scales, where the middle case represents the tradeoff point for the local bond ($\chi_c \cdot \chi_b = 256$).}
  \label{fig:profiles}
\end{figure*}

Two kinds of bottlenecks are considered. First, we examine the impact of ScaLAPACK routine calls like decomposition and matrix multiplication, comparing it to all other QTNH calls, including index permutation. The results are presented in Fig.~\ref{fig:profiles:la}, where it is clear that ScaLAPACK calls are the main bottleneck at all scales considered, exceeding 98\% at $\chi = 8192$. This is largely dominated by the decomposition calls, which encompass QR, pivoted QR and LQ factorisation routines, as well as the unitary generators \texttt{UNGQR}/\texttt{UNGLQ}. Hence, we confirm the decomposition bottleneck, even with our improved truncation algorithm. The GEMM calls also make up a significant fraction of the runtime, with the peak of $21\%$ at $\chi = 4096$. 


The second set of profiling diagrams, shown in Fig.~\ref{fig:profiles:op}, provides insight on the CPU/MPI contributions to the runtime. It should be noted that high communication costs are unavoidable in our approach due to poor scaling of the distributed factorisation calls~\cite{ferrero-roza-morinigo-23}. Indeed, for $\chi = 2048$, the emulation is highly communication-bound, with MPI taking over $70\%$ of runtime. However, this quickly changes as we increase the local size, with the $\chi = 4096$ showing a nearly perfect 50-50 CPU/MPI split. Moreover, this is a fine balance, since for $\chi = 8192$ not only is the run completely compute-bound, we also discover it is in fact more efficient to weakly scale the $\chi = 4096$ case to $8$~nodes. We believe those results can be explained by significantly longer memory access times on the rightmost plot, which is likely when the site tensors no longer fit in the L3~cache. This is further explored in Sec.~\ref{sec:results:riqft}, where we see a similar behaviour for the IQFT circuit. Ultimately, the crucial finding is that the most efficient local bond seems to be $\chi_c \cdot \chi_b = 256$, which is a long way before saturating the memory available on each node. 


\subsection{State-of-the-art comparison}
\label{sec:results:sota}

Even though the other libraries do not offer distributed implementations of the MPS evolution algorithm, we investigated the scaling of the local version, and how it compares to QTNH. Both ITensor and quimb were selected as the state of the art, since they offer an interface to MPS/MPO methods, and they are commonly used for circuit emulation~\cite{berezutskii-liu-25}. To test their limits, we executed the full version of the RCS circuit with increasing bond dimensions. The underlying system is dense due to the lack of quantum number preservation, and hence, the only parallelism available is BLAS threading. We found that both libraries run the fastest with \texttt{OMP\_NUM\_THREADS=16} when $\chi \geq 256$, but the speed-up factor never exceeded~$3$. 

Notably, we found running a comparison against CTF impractical, as it is a lower-level framework, and not commonly applied for quantum circuit emulation. As a result, realising our benchmarking circuits would prove a considerable software engineering work. We could instead have considered CTF as an alternative basis to implement our tensor-parallel MPS method, but we determined QTNH to be a better fit, since it is inherently aimed for dense tensor algebra and supports non-standard decomposition routines. 

The decomposition method prevalent in the state of the art is the SVD. Hence, we selected to compare it to both the SVD and pivoted QR algorithms in QTNH. We expect that QTNH with the SVD-based truncation is likely to give similar runtimes. Throughout the runs, we fixed $\chi^{(SVD)}_d = 8$, which is the largest power-of-two bond that fits on a single node. However, the pivoted QR method has inherently lower accuracy, and to make the comparison fair, we increased $\chi^{(pQR)}_d = 11$. This takes more time, but results in similar accuracy to the SVD approach, while still fitting within one node. 

Another consideration is how each library handles long-rage interactions. In quimb, the MPS is lazily permuted to bring the interacting qubits together -- this is an approach more robust than what we used in QTNH, since the sites are only moved on-demand. In ITensor, however, the interactions are always treated as MPOs, which is likely to be slow. Therefore, we used our bubble-sort permutation algorithm to make a more equal comparison. 

\begin{figure}[tb]
  \centering
  \includegraphics[width=\linewidth]{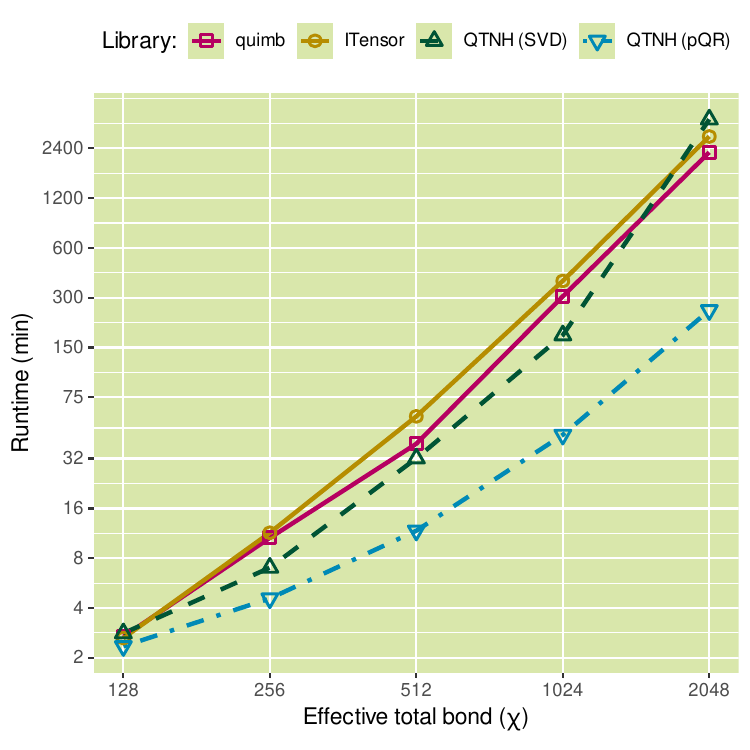}
  \caption{Comparison between the state-of-the-art libraries and QTNH, emulating the RCS circuit. Both ITensor and quimb were run with 16~BLAS threads, which was the fastest setting on ARCHER2. QTNH with SVD used the distributed bond $\chi_d = 8$ on 64~cores. QTNH with pivoted QR needed higher bond to achieve comparable fidelity, which was empirically set to $\chi_d = 11$. Lower runtime is better. }
  \label{fig:rcs-sota}
\end{figure}

The results are shown in Fig.~\ref{fig:rcs-sota}. We find that QTNH can outperform SOTA libraries for larger bonds, even on a single node. The difference is narrow for the SVD factorisation method, but it is clear that the pivoted QR approach runs faster even when compensating for the lower accuracy with an increased virtual bond. Neither quimb nor ITensor were able to reach $\chi = 4096$, which is the smallest bond that necessitates distribution in QTNH. The highest achievable bond dimension by SOTA was $\chi=2048$, which quimb computed the fastest, taking $38$~hours on ARCHER2, and resulting in fidelity of $\bar{\mathcal{F}} = 4.57 \cdot 10^{-19}$. In comparison, QTNH with the QR approach needed $\chi = 2816$ to achieve similar fidelity ($\bar{\mathcal{F}} = 6.81 \cdot 10^{-19}$), which only took $4.2$~hours on a single node, hence yielding $9\times$ speed-up. 


\subsection{Random IQFT}
\label{sec:results:riqft}

The IQFT initialised with a random state of fixed saturation is a perfect circuit to model moderate entanglement. This is because it doesn't introduce much entanglement on its own. Here, we show how the IQFT can be emulated with our tensor-parallel MPS to give nearly perfect fidelity. 


\subsubsection{Output fidelity}

\begin{figure}[tb]
  \centering
  \includegraphics[width=\linewidth]{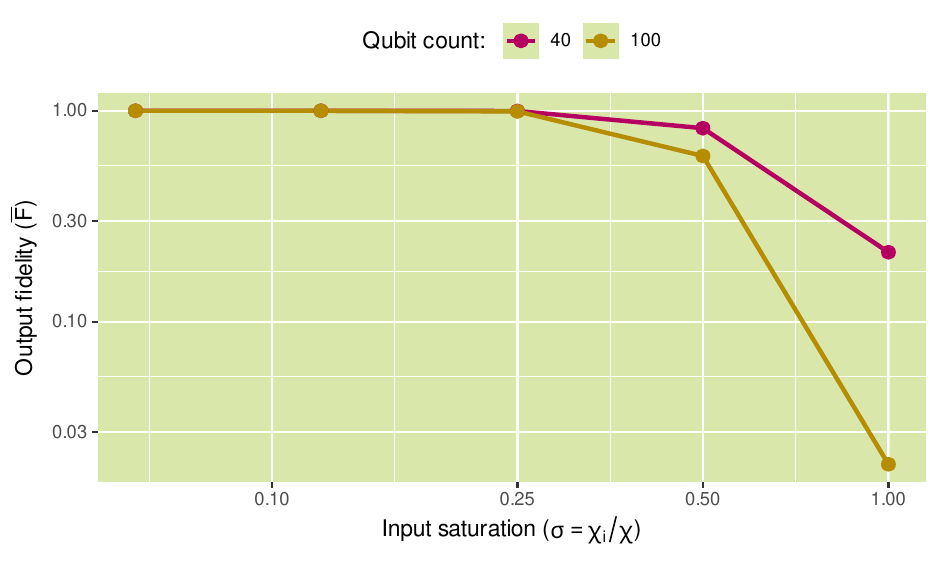}
  \caption{Fidelity of IQFT given fixed input bond saturation $\sigma=\chi_i/\chi$. Values of $\sigma \leq 0.25$ yield nearly perfect output. For larger systems, the fidelity drops faster with $\sigma$. }
  \label{fig:riqft-fidelity}
\end{figure}

When the input of IQFT is fully saturated, we expect the output fidelity to go down. In fact, Fig.~\ref{fig:riqft-fidelity} shows that the bigger the system, the higher the decoherence. However, at saturations of around $\frac{1}{4}$, the fidelity is already very close to $1$. This means, that if given a circuit which outputs a state with~$\chi=k$, we need~$\chi=4k$ to apply the IQFT with marginal error. 

\subsubsection{Strong scaling}

\begin{figure*}[tbp]
  \centering
  \includegraphics[width=.8\linewidth]{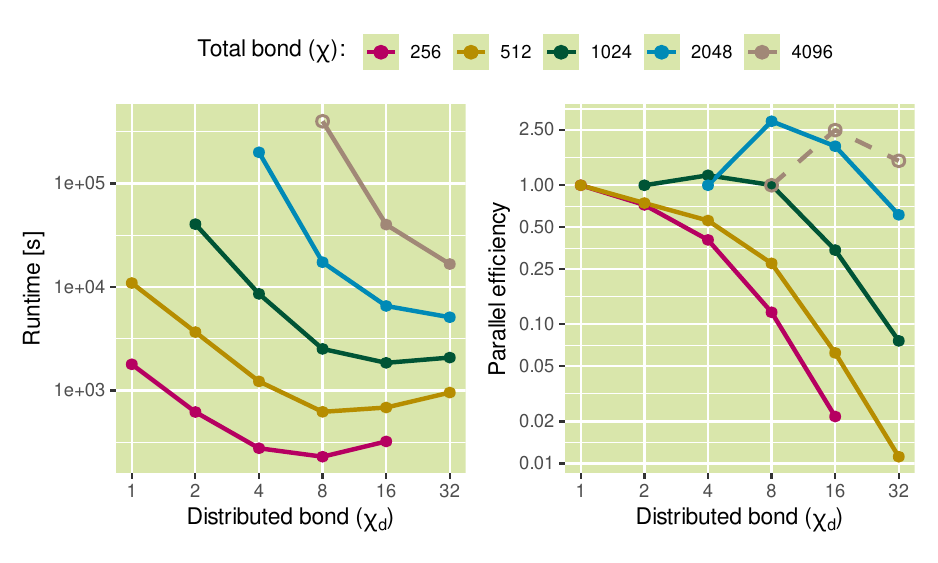}
  \caption{Strong scaling of $100$-qubit IQFT for fixed $\chi$. The number of ranks $r=\chi_d^2$. $\PE$ is relative to the minimum $r$ needed to fit the state ($\chi_l = 512$, $\leq 2$~GB per rank). \textbf{The key result is that saturating the ranks is inefficient}, as $\PE > 1$ when $\chi_l = 256$. This occurs at $2$~MB per local matrix, i.e. the maximum that fits into L3~cache. \\
  The empty dot on the left is from an incomplete run that exceeded $96$~hours (long QoS on ARCHER2). This has been extrapolated with high confidence based on the progress. Likewise, all the corresponding $\PE$ numbers on the right are also estimated. }
  \label{fig:riqft-strong}
\end{figure*}

The MPS-based methods are expected to scale poorly due to low parallel performance of decomposition methods~\cite{ferrero-roza-morinigo-23}. Nonetheless, we find optimal size per core for large-scale problems, which is demonstrated on Fig.~\ref{fig:riqft-strong}. Once the local bond $\chi_l = \chi_c \cdot \chi_b$ of the scaling baseline is $\chi_l = 512$, the strong PE actually exceeds~$1$. This is likely due to caching -- the memory required for each site at that scale is $\sim 8$~MB per core, and the size of L3~cache on ARCHER2 is $16$~MB per 4~cores. Therefore, the cache is too small to accommodate all matrix element, which likely hinders the performance. Moreover, at $100$~sites, the footprint of the entire MPS is over $800$~MB, or nearly half the available memory per core (without including ScaLAPACK work arrays). Hence, we conclude that $\chi_l=256$ is the best local size. 

\subsection{Google's circuit sampling}

RCS is an example of a circuit that rapidly maximises the entanglement. As a result, higher bond dimensions only marginally improve fidelity. The purpose of emulating such circuits is to benchmark the computational phase boundary between classical and quantum devices. In addition, this provides a reliable reference for comparing classical emulation software and algorithms. 


In this work, we took a straightforward strategy of MPS evolution, without more complex multi-qubit sites or the split-and-merge algorithm~\cite{zhou-stoudenmire-20}. Our approach is scalable and can be universally applied to other circuits, but the trade-off is that it is not specifically optimised for the RCS algorithm. Hence, our aim is not to reach the fidelity of Sycamore, but instead to demonstrate the improved scaling of a problem that is classically-hard to simulate. 

\subsubsection{Weak scaling}

\begin{figure*}[tbp]
  \centering
  \includegraphics[width=.8\linewidth]{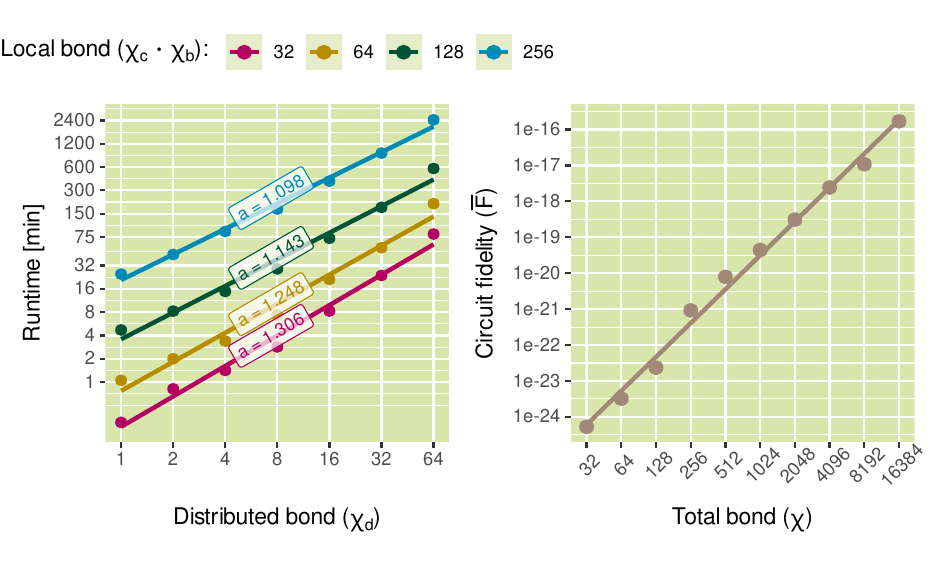}
  \caption{Weak scaling of Google's 53-qubit RCS emulation with SWAP + QR approach, at fixed local $\chi_l$. The results are shown on the log-log plot, as the underlying numerical method dictate at least linear scaling, which corresponds to the gradient of $1$. The actual gradients are annotated, and they indicate the actual performance of the distribution -- the closer to $1$, the better. Plot to the right shows output fidelity (independent of distribution). At $\chi=16,384$, we achieve $\bar{\mathcal{F}} = 1.69 \cdot 10^{-16}$, over $10^8$ times better than for $\chi=32$. }
  \label{fig:rcs-weak}
\end{figure*}

When increasing the distributed size of MPS, we need to account for the complexity of the underpinning numerical approach, which is close to cubic for both decomposition and matrix multiplication. As we double each matrix dimension, the number of elements quadruples, so to evaluate weak scaling, we need to distribute them to four times the number of processes. However, the number of arithmetic operations is $\mathcal{O}(n^3)$, and the scaled problem would need 8~times as many. Therefore, the expected runtime is doubled even with no additional communication costs, and it should increase linearly with the distributed bond dimension ($\chi_d$). In Fig.~\ref{fig:rcs-weak}, we see experimental results match this expectation, with an optimal log-log plot gradient of~$1.1$, which indicates that our implementation features marginal overheads beyond the underlying numerical methods. 



\subsubsection{Output fidelity}

Our largest experiment was a $\chi=16,384$ MPS, which took $40.8$~hours on $32$~ARCHER2 nodes to complete for the entire RCS circuit. We managed to achieve the fidelity of $\bar{\mathcal{F}} = 1.69 \cdot 10^{-16}$, or $\bar{f}_{av} = 91.9\%$, and $\bar{f}_{swap} \approx 98.0\%$. 


As described in Sec.~\ref{sec:results:sota}, the best accuracy of the state-of-the-art libraries was $\bar{\mathcal{F}} = 4.57 \cdot 10^{-19}$ for the bond dimension of $\chi=2048$, computed in at least $38$~hours on a single node. Since the other libraries do not allow distributed execution, this was the largest problem size achievable on ARCHER2. Hence, we managed to attain $8\cdot\chi$ of the state of the art, improving the accuracy $370$~times in a comparable time by scaling to $32$~nodes of ARCHER2.








\section{Future work}
\label{sec:future}

We have presented a novel approach to distribute matrix product states for quantum circuit emulation, which is implemented under a brand new C++ library. This leaves room for a number of optimisations and improvements that didn't fit in the scope of the work. 

\begin{enumerate}
  \item \textbf{Integration with all types of tensor parallelism} -- the library currently supports tensor parallelism with distribution pattern directly compatible with network parallelism. However, the latter can't be used due to limitations of ScaLAPACK which uses globally blocking MPI communicators. If this gets corrected, tensor network distribution will follow naturally from the offset parameter. Slicing parallelism can also be integrated, but we believe it wouldn't be of particular help for representations with decomposition bottlenecks. 
  \item \textbf{Accelerator support} -- since we reached the scale where our method is CPU-bound, an obvious improvement would be offloading computations to accelerators like GPUs. However, one of the biggest missing pieces for distributed GPU-accelerated MPS are decomposition methods. In this work, we showed that the SVD can be replaced with pivoted QR, which could be a simpler option to enable accelerator support. Alternatively, there is a study where polar decomposition was shown to be another option, implemented on tensor processing units~\cite{ganahl-beall-23}. 
  \item \textbf{More complex representations} -- we were unable to outperform the supremacy circuit using a simple MPS algorithm. However, there are approaches which can get closer to that target at much lower computational costs. We believe that QTNH with our distribution pattern would be a perfect setting to implement them at a larger scale. 
  \item \textbf{Other MPS-based techniques} -- the original purpose of MPS is not the emulation of quantum circuits, but rather more general physics problems via density matrix renormalisation group (DMRG) or time-evolving block decimation (TEBD) algorithms. The block-cyclic MPS should enable straightforward parallelisation of other similar methods. 
\end{enumerate}

\section{Conclusions}
\label{sec:conclusions}

We introduced a novel distribution pattern for the MPS quantum circuit evolution algorithm, which has potential to increase the scale and accuracy of classical emulation of quantum circuits. This was achieved by leveraging tensor parallelism, and treating individual site tensors as evenly block-distributed matrices, which helps evade the problems arising from the sequential nature of MPS recanonicalisation. Furthermore, our approach is naturally load-balanced and scalable up to the limits of the underpinning numerical methods, such as matrix decomposition. Hence, we offer a new method to determine the computational phase boundary between classical and quantum devices. 

Implementing the tensor-parallel MPS algorithm is possible thanks to our QTNH library, which addresses the gap in distributed dense tensor software. It offloads the operations like tensor contraction and decomposition to ScaLAPACK, enabling high-performance applications. We find that one of the major bottlenecks of MPS-based methods is singular value decomposition (SVD) needed for tensor truncation. This is addressed by employing a different type of sorting decomposition, pivoted QR factorisation, which offers significantly better runtimes at the price of result fidelity. We apply it to emulate the widely known random circuit sampling~(RCS) benchmark, and manage to improve the single-node runtime of the state-of-the-art methods by up to $9$~times for the same accuracy, despite larger system size. 

We also leverage our distribution model to run the RCS benchmark with unprecedented dense MPS evolution bond dimension of $\chi=16,384$, using $32$~nodes of ARCHER2, and taking $40.8$~hours of runtime. In contrast, none of the other libraries supports multi-node runs, and the best result was from the quimb Python package, which took $37.8$~hours to reach $\chi=2048$. Hence, our method yields a $370\times$ improvement in the output fidelity. 



This study contributes to the field of quantum emulation by distributing the MPS evolution scheme, which enables higher-accuracy simulation of larger, more entangling quantum circuits than what can be achieved using only shared memory. This is particularly pertinent for dynamical simulations such as quantum circuit emulation, as once the bond dimension is saturated accuracy is lost, and cannot be recovered. Furthermore, our tensor-parallel approach is compatible with more complex representations and algorithms, and we hope that QTNH can make further impact by unlocking other large-scale tensor network implementations. 

\section*{Acknowledgements}

The authors thank Nick Brown for helpful discussions and very useful feedback given during the preparation of this manuscript. This work used the ARCHER2 UK National Supercomputing Service (\url{https://www.archer2.ac.uk}) \cite{ARCHER2}.

Funding: JA acknowledges the support of EPSRC studentship 2708175.

\bibliographystyle{elsarticle-num} 
\bibliography{refs}

\appendix

\section{Quantum circuits}
\label{sec:app:qc}


\begin{figure}[tbp]
  \centering
  \includegraphics[width=\linewidth]{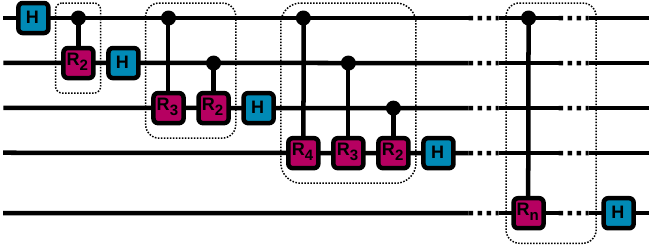}
  \caption{A general IQFT circuit diagram. Swaps are omitted, as they only affect the ordering of the result, and instead the input qubits are reversed. The controlled phase shifts with the same target can be grouped together into a single gate, which is more efficient to use in the emulation. }
  \label{fig:iqft-circuit}
\end{figure}

\begin{figure}[tbp]
  \centering
  \includegraphics[width=\linewidth]{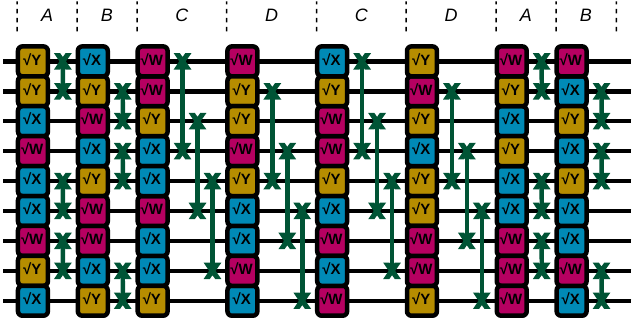}
  \caption{An example RCS circuit diagram of a $3\times3$ qubit grid. As the diagram orders the qubits in one dimension, not all interacting qubits appear next to each other, even though they are nearest neighbours in 2 dimensions. The actual RCS circuits used in this work is much larger, and orders the qubits as shown on Fig.~\ref{fig:rcs-patterns}. }
  \label{fig:rcs-circuit}
\end{figure}

For reference, here we outline the implementation details of the circuits used in this work. Fig.~\ref{fig:iqft-circuit} shows a general \textbf{Inverse Quantum Fourier Transform} circuit. The Hadamard gate $\hat{H}$ and phase shift gate $\hat{R}_k$ are defined as follows: 

\begin{equation}
  \hat{H} = \frac{1}{\sqrt{2}} \begin{pmatrix}
    1 & 1 \\
    1 & -1
  \end{pmatrix} \qquad 
  \hat{R}_k = \begin{pmatrix}
    1 & 0 \\
    0 & e^{2\pi i / 2^k}
  \end{pmatrix} 
\end{equation}

For improved efficiency, multiple controlled $\hat{R}_k$ gates on the same target can be applied as a single gate. We also omit the swap gates at the beginning or end of the circuit, as they only rearrange the qubits in the result. 

Fig.~\ref{fig:rcs-circuit} visualises how the \textbf{Random Circuit Sampling} is implemented. It involves the following three gates applied at random: 

\begin{equation}
  \begin{gathered}
    \hat{X}^{1/2} = \frac{1}{\sqrt{2}} \begin{pmatrix}
      1 & -i \\
      -i & 1
    \end{pmatrix} \qquad 
    \hat{Y}^{1/2} = \frac{1}{\sqrt{2}} \begin{pmatrix}
      1 & -1 \\
      1 & 1
    \end{pmatrix} \\
    \hat{W}^{1/2} = \frac{1}{\sqrt{2}} \begin{pmatrix}
      1 & -\frac{1 + i}{\sqrt{2}} \\
      \frac{1 - i}{\sqrt{2}} & 1
    \end{pmatrix}
  \end{gathered}
\end{equation}

\begin{figure*}[tp]
  \centering
  \subfloat[]{
    \includegraphics[width=0.4\linewidth]{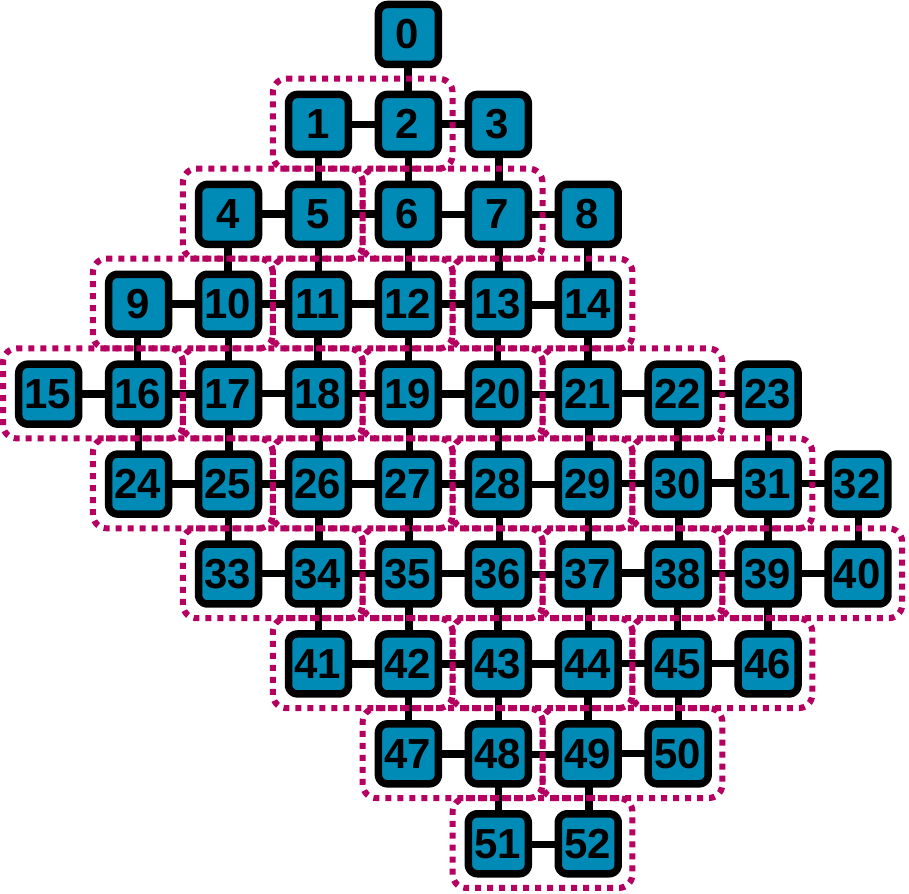}
    \label{fig:rcs-pattern-a}
  } \qquad
  \subfloat[]{
    \includegraphics[width=0.4\linewidth]{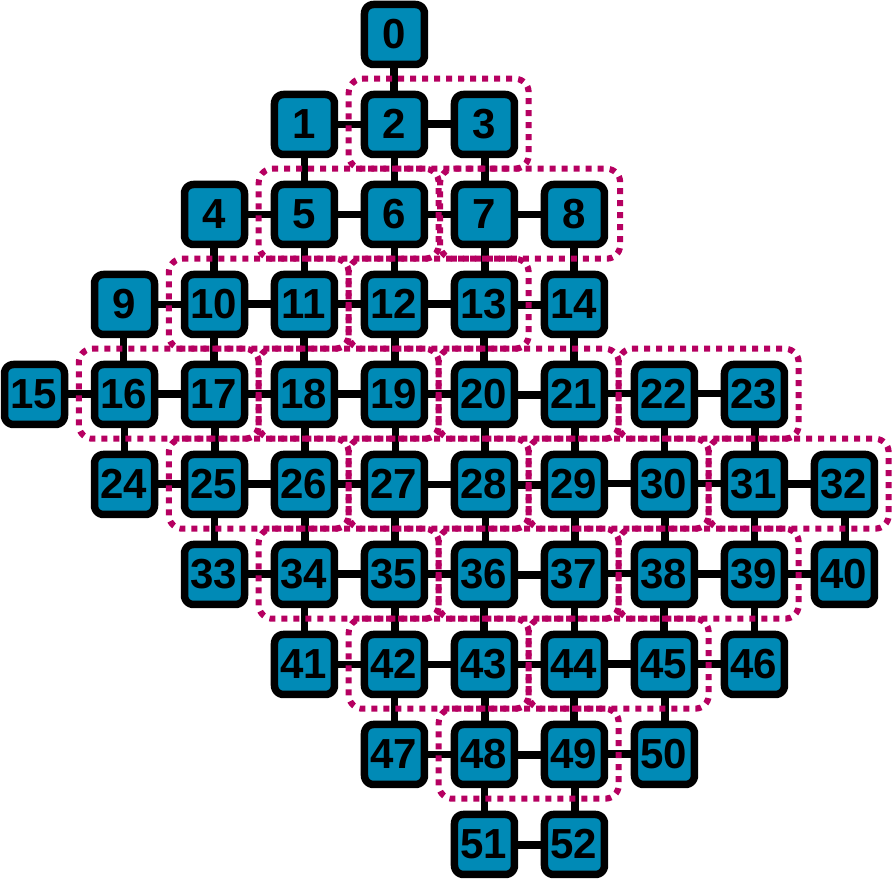}
    \label{fig:rcs-pattern-b}
  } \\
  \subfloat[]{
    \includegraphics[width=0.4\linewidth]{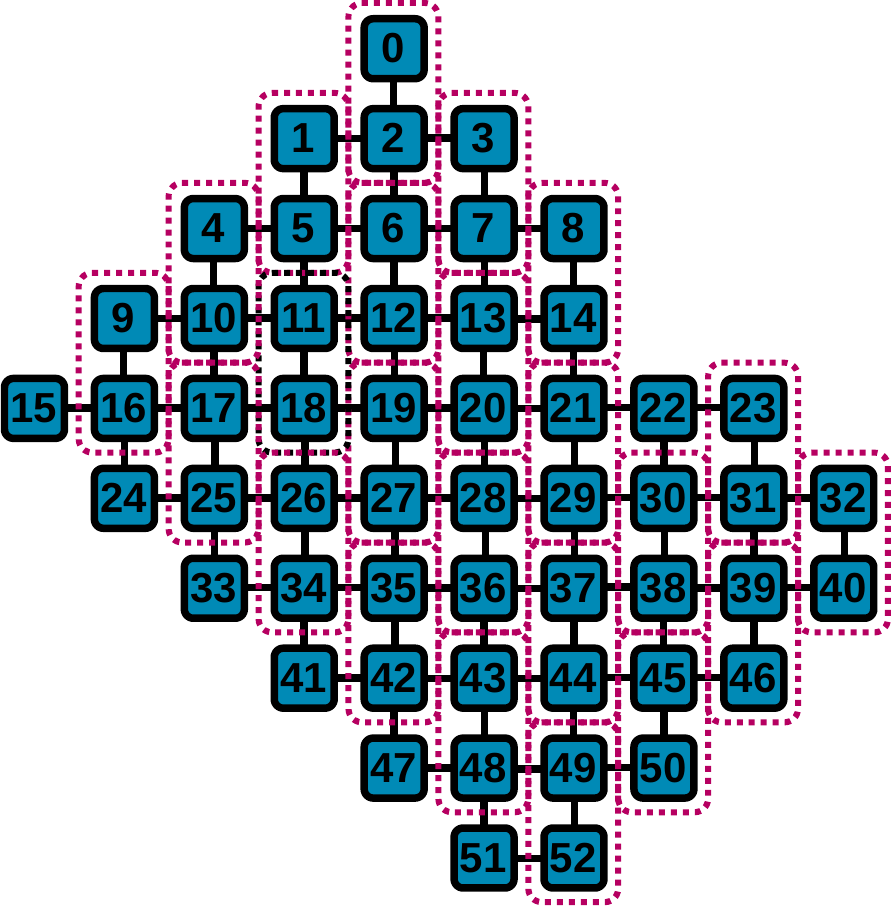}
    \label{fig:rcs-pattern-c}
  } \qquad
  \subfloat[]{
    \includegraphics[width=0.4\linewidth]{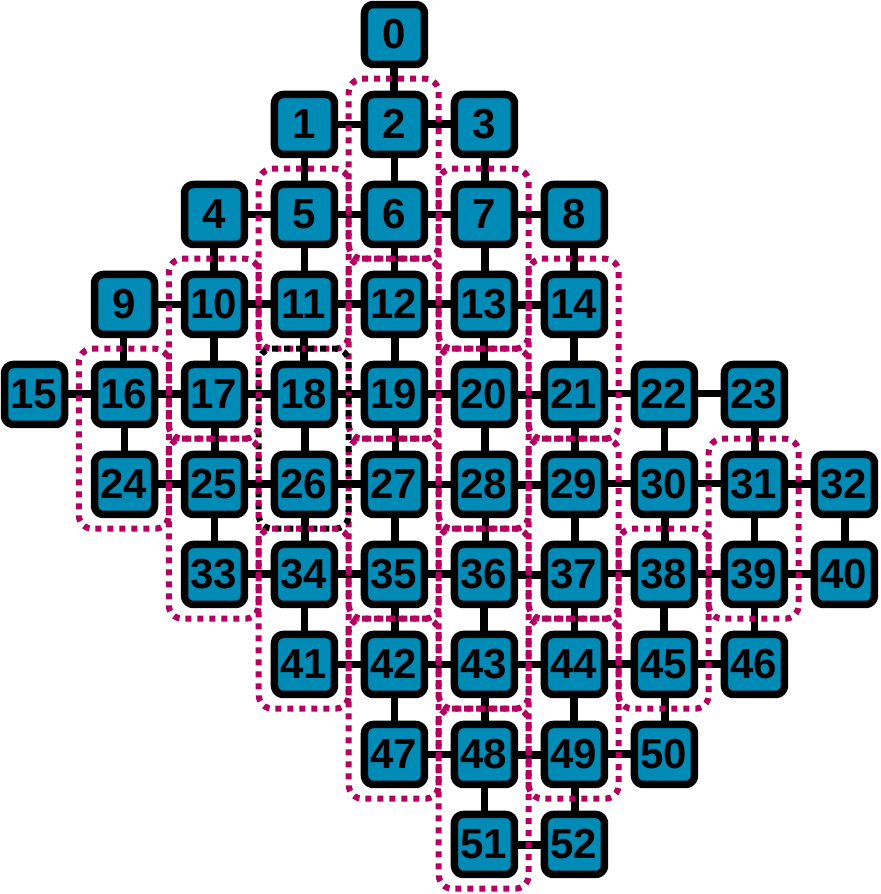}
    \label{fig:rcs-pattern-d}
  }

  \caption{RCS topology and interaction patterns, including the initial labelling of qubits. Patterns $C$ and $D$ are long-range in one dimension, and therefore more difficult to apply. }
  \label{fig:rcs-patterns}
\end{figure*}

The entangling $\fSim$ gate (shown in green on the diagram) is presented in Eqn.~\ref{eqn:fsim}. The targets of the entangling layers are the nearest neighbours on the two-dimensional Sycamore topology, according to one of four patterns ($A$, $B$, $C$, $D$), as demonstrated on Fig.~\ref{fig:rcs-patterns}. The RCS circuits applies them in the order $ABCDCDAD$ to ensure high entanglement.







\end{document}